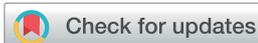

## PAPER

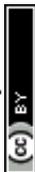

Check for updates

# Influence of the electrode nano/microstructure on the electrochemical properties of graphite in aluminum batteries†

Cite this: *J. Mater. Chem. A*, 2018, **6**, 22673

Giorgia Greco, [ID] *[a] Dragomir Tatchev, [ID][b] Armin Hoell, [ID][a] Michael Krumrey, [ID][c] Simone Raoux, [ID][ad] Robert Hahn[ef] and Giuseppe Antonio Elia [ID] *[f]

Herein we report on a detailed investigation of the irreversible capacity in the first cycle of pyrolytic graphite electrodes in aluminum batteries employing 1-ethyl-3-methylimidazolium chloride:aluminum trichloride (EMIMCl:AlCl$_3$) as electrolyte. The reaction mechanism, involving the intercalation of AlCl$_4^-$ in graphite, has been fully characterized by correlating the micro/nanostructural modification to the electrochemical performance. To achieve this aim a combination of X-ray diffraction (XRD), small angle X-ray scattering (SAXS) and computed tomography (CT) has been used. The reported results evidence that the irreversibility is caused by a very large decrease in the porosity, which consequently leads to microstructural changes resulting in the trapping of ions in the graphite. A powerful characterization methodology is established, which can also be applied more generally to carbon-based energy-related materials.

## Introduction

Lithium-ion batteries are the power source of choice for portable electronic devices and for hybrid and full electric vehicles.[1–3] However, the use of lithium in the main power source for electromobility is questionable due to limited lithium resources.[4] Alternatives to lithium, such as Na,[5–10] K,[11–13] Ca,[14,15] Mg[16,17] and Al,[18–20] have attracted increasing interest due to the greater abundance of these elements. Aluminum is the most abundant metal element in the Earth's crust and is characterized by an extremely high volumetric capacity of 8040 mA h cm$^{-3}$, which is four times higher than that of lithium. It also has a satisfactory specific capacity of 2980 mA h g$^{-1}$, making it an extremely good candidate for the realization of alternative electrochemical storage systems.[18] The most advanced battery system employing an aluminum anode takes advantage of the anion intercalation process in graphite cathodes.[19,21–31] In a previous paper[25] we have investigated the reaction mechanism of the aluminum/graphite system employing a pyrolytic graphite electrode, evidencing the intercalation of the AlCl$_4^-$ anions in-between graphite layers, and showed an excellent stability for more than 2000 cycles in that system. On the other hand, we have also reported an irreversibility of 30% in the first (dis-)charge cycle showing that this phenomenon is most likely associated with the partial retention of anions intercalated between the graphite layers.[25] To understand the cause of this irreversibility, in the present work we report a thorough investigation of the mechanism of AlCl$_4^-$ intercalation into pyrolytic graphite in aluminum batteries by combining *ex situ* X-ray diffraction (XRD), small-angle X-ray scattering (SAXS) measurements, and computed tomography (CT). The implementation of these three techniques allows comprehensive investigation on different length scales, from nano to micrometers, respectively, resulting in a detailed understanding of the intercalation process.

## Experimental

The electrolyte 1-ethyl-3-methylimidazolium chloride:aluminum trichloride (EMIMCl:AlCl$_3$) in a 1 : 1.5 molar ratio was provided by Solvionic, with the water content of the electrolyte being lower than 100 ppm. The electrochemical measurements were performed using Teflon Swagelok® type T cells.[20,25] All potentials quoted in this manuscript are referenced to the quasi reference Al/Al$^{3+}$ electrode. Pyrolytic graphite (PG, Panasonic EYGS121810) with a thickness of 100 ± 30 μm, a loading of 8.66 mg cm$^{-2}$ and an average density of 0.85 g cm$^{-3}$ was used as the working electrode. The initial thickness of the electrodes was chosen to be 100 ± 2 μm. The galvanostatic cycling tests of Al/EMIMCl:AlCl₃/PG cells were carried out by applying a specific current of 25 mA g⁻¹ in the voltage range from 0.4 to 2.4 V using a 10 mm diameter disk of PG as the working electrode and a 10 mm diameter disk of Al (99.99% Alfa Aesar) as the counter electrode. All specific currents (mA g⁻¹) and specific capacities (mA h g⁻¹) mentioned in the manuscript are referenced to the cathode (PG) mass. For *ex situ* characterization, cells were stopped at different degrees of intercalation during the first cycle (see Fig. 1a): after a charge of 25 mA h g⁻¹ (Ch25, green), after full charge (FullCh, red), after a discharge of 50 mA h g⁻¹ (Dis50, orange), and after full discharge (FullDis, yellow). After disassembling the cell, the electrodes were rinsed for 30 s in dimethyl carbonate (DMC, Sigma-Aldrich, 99.9% anhydrous) to remove any residual electrolyte.[25] In order to avoid reaction with oxygen or moisture, cell preparation and disassembly, as well as sample preparation and rinsing, were performed in an argon filled glove box with oxygen and water concentrations lower than 0.1 ppm. All the transfers from the glove box to the measurement instruments were performed using airtight containers.

[a]*Helmholtz-Zentrum Berlin für Materialien und Energie GmbH, Hahn-Meitner-Platz 1, D-14109 Berlin, Germany. E-mail: giorgia.greco@helmholtz-berlin.de*

[b]*Institute of Physical Chemistry, Bulgarian Academy of Sciences, Acad. G. Bonchev Str. Bl.11, 1113 Sofia, Bulgaria*

[c]*Physikalisch-Technische Bundesanstalt, Abbestr. 2-12, D-10587 Berlin, Germany*

[d]*Institut für Physik, Humboldt-Universität zu Berlin, Newtonstr. 15, D-12489 Berlin, Germany*

[e]*Fraunhofer-Institut für Zuverlässigkeit und Mikrointegration, Gustav-Meyer-Allee 25, D-13355 Berlin, Germany*

[f]*Technische Universität Berlin, Research Center of Microperipheric Technologies, Gustav-Meyer-Allee 25, D-13355 Berlin, Germany. E-mail: elia@tu-berlin.de*

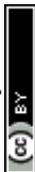

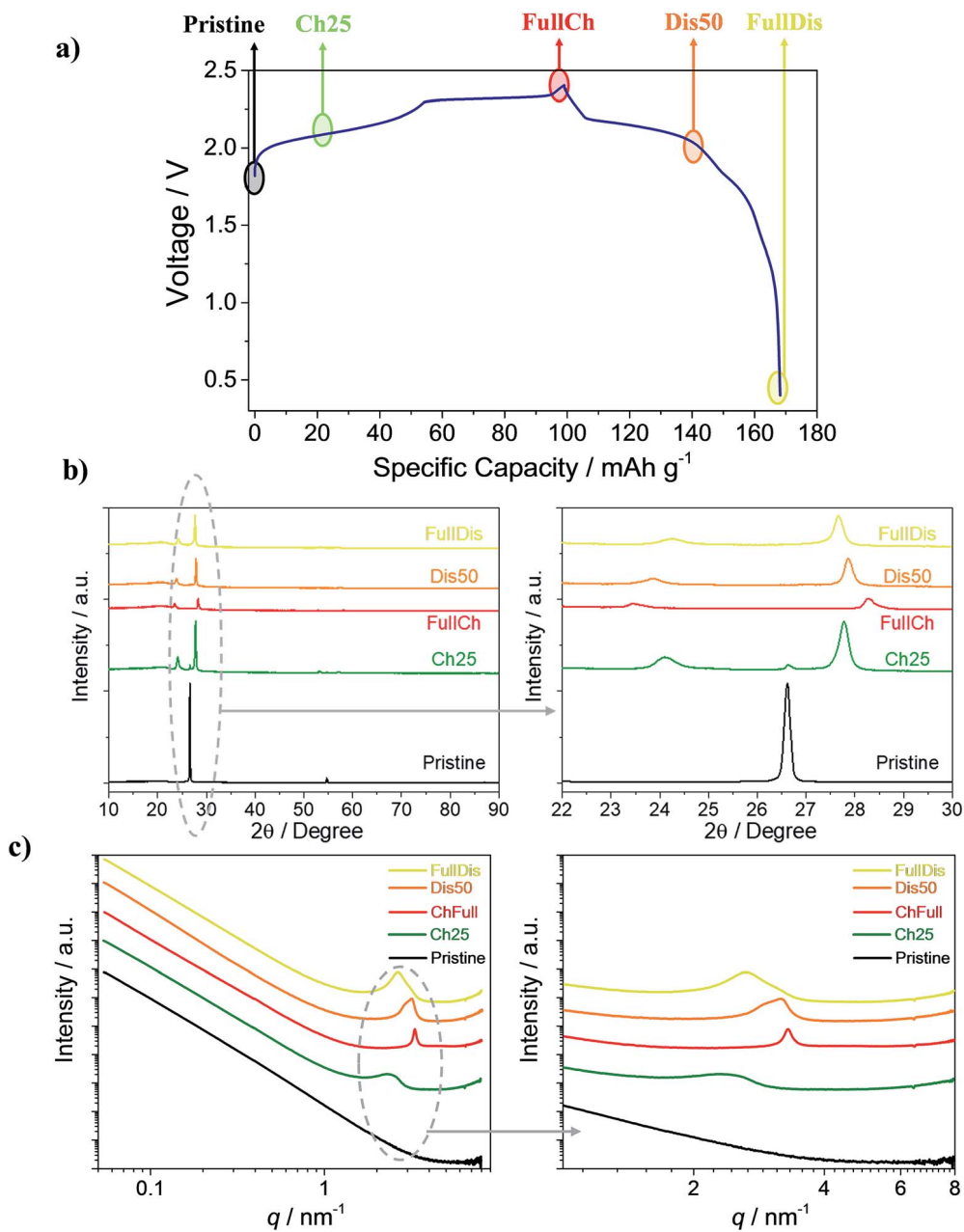

Fig. 1 (a) Voltage profile of the Al/EMIMCl:AlCl₃/PG cell galvanostatically cycled at a current of 25 mA g⁻¹ at 25 °C in the 0.4 to 2.4 V cut-off voltage range highlighting the charge states of the PG used for the *ex situ* measurements. (b) XRD diffractograms of PG electrodes at different charge states. (c) Small-angle X-ray scattering curves at 10 keV of the PG electrodes at different charge states. Charged 25 mA h g⁻¹ (Ch25 – green), fully charged (FullCh – red), discharged 50 mA h g⁻¹ (Dis50 – orange) and fully discharged (FullDis – yellow).

The X-ray diffraction (XRD) measurements of the prepared electrodes were performed using Cu Kα radiation ($\lambda = 0.154$ nm) on a Bruker D8 Advance diffractometer in the $2\theta$ range from $10°$ to $90°$ with a step size of $0.01°$. The electrodes were fixed on the diffractometer's sample holder in the glove box using Kapton tape to isolate the electrode from the atmosphere during measurements. The SAXS measurements were carried out at the beamline FCM (four crystal monochromator)[32,33] of the Physikalisch-Technische Bundesanstalt (PTB) at the synchrotron radiation facility BESSY II of Helmholtz-Zentrum Berlin (HZB). The beamline was equipped with an anomalous SAXS (ASAXS) instrument designed and constructed by HZB.[34] The primary-beam cross-section at the sample position was 0.5 mm × 0.5 mm, the used sample–detector distances were 3721 mm and 1426 mm, respectively, and the photon energy was 10 keV, equivalent to a wavelength $\lambda = 0.124$ nm. The detector used for SAXS is a vacuum-compatible PILATUS 1M.[35] Scattering patterns were recorded with an exposure time of 25 min. The obtained patterns cover the region of $0.05$ nm$^{-1} \leq q \leq 4.2$ nm$^{-1}$ where $q$ is defined as $q = (4\pi/\lambda)\sin(\theta/2)$, where $\theta$ is the scattering angle. All SAXS sample measurements were circularly averaged and corrected for sample transmission and background. For each measurement a sample of Ag-behenate powder was measured as a standard for the $q$-axis calibration.[36]

The tomography measurement was carried out with a microtomograph, Bruker SkyScan 1272, that uses a white beam with a cone geometry. The following set-up conditions were applied: an X-ray tube voltage of 80 kV, a current of 124 μA, and a filter of 1 mm Al. The voxel (3D pixel) size was 1 μm and the optical magnification was 7.4. Strips of about 4 mm width were cut from 1 cm diameter Kapton encapsulated PG electrodes. The strips were tightly wrapped in polyethylene foil to avoid prolonged air exposure. Five samples were packed and scanned together. The 360° scan lasted 13 hours and 14 minutes. The reconstruction of 3D images was done with the commercial software InstaRecon. The samples were analyzed separately one by one by manually selecting the volume of interest (VOI) entirely within the analyzed electrode. Damaged areas of the electrode were excluded. The Bruker proprietary software CTAn was used for 3D segmentation and determination of VOI histograms, porosities, structure thickness and structure separation of the electrodes.

## Results and discussion

Fig. 1a shows the voltage profile of the Al/EMIMCl:AlCl$_3$/PG cell galvanostatically cycled at 25 mA g$^{-1}$. The voltage profile follows the expected multi-plateau shape, which is typical of anion intercalation in graphite,[19,21,25–27,37–40] with a specific charge capacity of 100 mA h g$^{-1}$ and a discharge of 70 mA h g$^{-1}$, leading to a first cycle coulombic efficiency of 70%. In the following cycles the coulombic efficiency increases to values higher than 98% and the cells reversibly deliver 70 mA h g$^{-1}$, as already reported in a previous paper.[25] To understand the reasons behind the partial retention of AlCl$_4^-$ in PG, samples with various degrees of intercalation, namely the pristine sample, the PG charged to 25 mA h g$^{-1}$ (Ch25), the fully charged PG (FullCh), the PG discharged to 50 mA h g$^{-1}$ (Dis50), and the fully discharged PG (FullDis), were investigated using XRD, SAXS and CT.

During the intercalation process a new material called a graphite intercalated compound (GIC) is formed. A scheme of the GIC is reported in Fig. S1,† highlighting the GIC's characteristic parameters such as the periodic repeat distance $I_c$ and the gallery height $d_i$.

Fig. 1b shows *ex situ* X-ray diffractograms of the investigated samples, with magnification in the $22°$ to $30°$ range. Upon charging, the graphite (002) reflection disappears and two new peaks appear, indicating the formation of a GIC. The most intense reflection is associated with the $(00n + 1)$ periodic arrays and the second most intense one is associated with the $(00n + 2)$ periodic arrays of the intercalated layers between the graphene planes.

From the positions of the two most intense reflections, the so-called intercalation stage, $n$, representing the number of free graphene planes between two anion intercalated planes, can be obtained. The stage, $n$, can be obtained using theoretical calculation tabulated elsewhere[46,50] (see also Fig. S1†).

Eqn (1):[40–43]

$$I_c = l \, d_{obs} \tag{1}$$

can be used to evaluate the periodic distance ($I_c$) from the observed spacing, $d_{obs} = \lambda/2 \sin(2\theta)$, obtained from the main peak position $2\theta$, wavelength $\lambda$ and its Miller index $l(00n + 1)$ associated with the intercalation stage. The reported $I_c$ is the average resulting from the two main peaks and for each region in the charge/discharge cycle in Table S1.†

Since the repeat distance is on the order of nanometers, the corresponding interference peak appears in the small-angle scattering range and its wave vector position, $q_c$, is inversely proportional to the periodic distance $I_c$ given by eqn (2):

$$q_c = 2\pi/I_c. \tag{2}$$

Fig. 1c shows the SAXS curves obtained from the PG samples with various degrees of intercalation reported on a double logarithmic scale.

The graph shows the appearance of a broad peak upon a charging of 25 mA h g$^{-1}$ (green line). This peak shifts to higher $q$-values and becomes much sharper for the fully charged state. Upon discharge (de-intercalation) the peak shifts to lower $q$-values and becomes broader. The formation of the peak indicates the formation of a regular electron density distribution in the sample, associated with the stage of intercalation of the graphite cathode.[25] To evaluate the reported results, the scattering curves were fit by a combination of a power function (eqn (S1)†), for fitting the initial part of the curve ($0.06$ nm$^{-1} \leq q \leq 1$ nm$^{-1}$), and a Voigt function (eqn (S2)†), for fitting the peak ($1$ nm$^{-1} \leq q \leq 4$ nm$^{-1}$). The fitting results are shown in Table S2 in the ESI† section; a plot of the curve for the fully charged PG electrode (red line) and the fitted curve (dotted gray line) are reported in Fig. S2.†

The SAXS pattern of the pristine PG is typical for a certain class of carbon materials.[44] Porous graphite was reported to satisfy the Porod law, i.e. the scattering intensity decreases with $q^{-4}$, at least for point collimation.[45] For the investigated PG the obtained exponent varies slightly between 3.70 and 3.54 and does not show particular changes upon cycling (Table S2†). The average slope at high $q$ is a bit smaller than 4, namely $3.62 \pm 0.07$. Such values of the power law exponent are also typical for carbon materials.[44] Three explanations are possible: density fluctuations within the carbon sheets,[46,47] fractal structure,[48,49] and the specific shape of the particle size distribution.[49,50]

However, as the data can be fitted with a superposition of the contributions given in eqn (S1) and (S2),† the scattering contribution of the PG itself does not disturb the determination of the peak positions. The Voigt function fitting values of the peak in the scattering curves are reported in Table S2.†

The SAXS curves of the PG Ch25, FullCh and FullDis samples well fit with a single peak while the curve of the Dis50 sample fits well with two peaks, suggesting the presence of two different GIC phases in the Dis50 sample. From the obtained results we can evaluate the intercalant gallery expansion $\Delta d$ using formula (3):

$$\Delta d = I_c - 0.335\, n = d_i - 0.335\, (n + 1) \tag{3}$$

where $n$ is the intercalation stage obtained using XRD, $I_c$ is the periodic repeat distance measured using SAXS and 0.335 nm is the lattice parameter of graphite.[19,23,25]

The value of $\Delta d$ should be independent with respect to the intercalation stage, as it is mainly correlated to the dimension of the intercalated species.[51] The obtained gallery expansion, $\Delta d$, values for different samples are shown in Table S3.†

Table S3† summarizes the values of the periodic repeat distances ($I_c$) and dominant stages ($n$) calculated from XRD and SAXS as reported in Fig. 1b and c, respectively, and shows a good agreement between the results obtained from the two techniques. The obtained values of the intercalation gallery expansion, $\Delta d$, are very close to previously reported values for the intercalation of chlorine compounds in graphite[51] and within the uncertainty coincide with our previous measurements.[25]

The reported results indicate that the electrochemical intercalation leads to the formation of a stage 6 intercalated compound in the initial stage, and finally forms a stage 4 compound at full charge. During the de-intercalation process an intermediate stage 5 intercalated compound is formed and at full discharge only stage 6 is present. The fully discharged stage differs from the previously evaluated stage 7,[25] and could not be identified in our previous estimation because of an inability to distinguish stages 6 and 7. The combination with the SAXS results allowed us to more accurately indicate a dominant stage 6. Moreover, the width and the area of the SAXS peak give us information about the ordering and quantity of the intercalated species, respectively. For the fully charged electrode the sharp peak reveals a well-defined stage over the entire cathode, while the fully discharged electrode shows a broader peak that reveals a more disordered distribution of the species (see also Fig. 1 and Table S2†).

Moreover, the reported results clearly evidence irreversibility in the first cycle due to the retention of intercalated species in-between the graphite layers.[25] One reason for this behavior could be associated with the presence of functional groups or defects in the graphite that can chain some of the anions intercalated during the first cycle. However, the employed pyrolytic graphite is extremely pure and crystalline, excluding that hypothesis. A second possibility is that the irreversible capacity is associated with the structural variation of the electrode upon cycling.

To investigate how the intercalation process and the formation of GICs affect the microstructure of the electrode, an ex situ tomography measurement was carried out.

Typical cross-sections of the four charged electrodes and the pristine PG are shown in Fig. 2a. The PG shows a porous structure that tends to be layered with layers in the plane of the graphite sheet. The electrode thickness increases from the nominal 100 μm of the pristine PG to 144 μm in the fully charged state, see Table S4† and Fig. 2c. The in-plane cross-sections reported in Fig. 2b reveal no visible difference in the electrode morphology, while a clear change in the brightness is evidenced, which is also noticeable in Fig. 2a, suggesting increased X-ray absorption that is usually related to an increased mass density.

Segmentation, i.e. the conversion of grayscale 3D images to black and white images, is a major source of inaccuracy in computed tomography. Thus, aside from the analysis of the black and white 3D images, an analysis of the histograms was also done. The grayscale index histogram of a 3D image of an ideal k-phase system would comprise only k non-zero values, corresponding to the k phases. The histogram of a 3D image obtained by CT scanning contains peaks rather than single values corresponding to phases. The peak shape is often Gaussian, as in the case studied here.

The segmentation of the 3D images was done using the Otsu method.[52] The grayscale value of the threshold is shown in Table S4.† The porosity calculated after segmentation is also depicted in Fig. 4. The structure thickness and structure separation were also determined and are presented in Fig. S3.†

The structure thickness is the mean of the structure thickness distribution. The latter represents the diameter distribution of a system of spheres entirely contained in the structure, but encompassing each point of it.[53] When applied to the matrix, i.e. the voids, the structure thickness is called structure separation since in this case it is a measure of the distances between the objects in the structure. Due to its nature, the structure thickness tends to give the average of the smallest size in a random structure. Since the PG electrodes show an approximately layered structure, the structure thickness corresponds to the thickness of the carbon layers, while the structure separation corresponds to the average distance between them. Finally, the degree of anisotropy and the fractal dimension were determined and are shown in Table S4.†

Fig. 3 shows the histograms for the five electrodes under study. They decompose very well into two Gaussian peaks. The peak at the lower grayscale index corresponds to the phase with the lower absorbing power, i.e. the voids. The peak at the higher

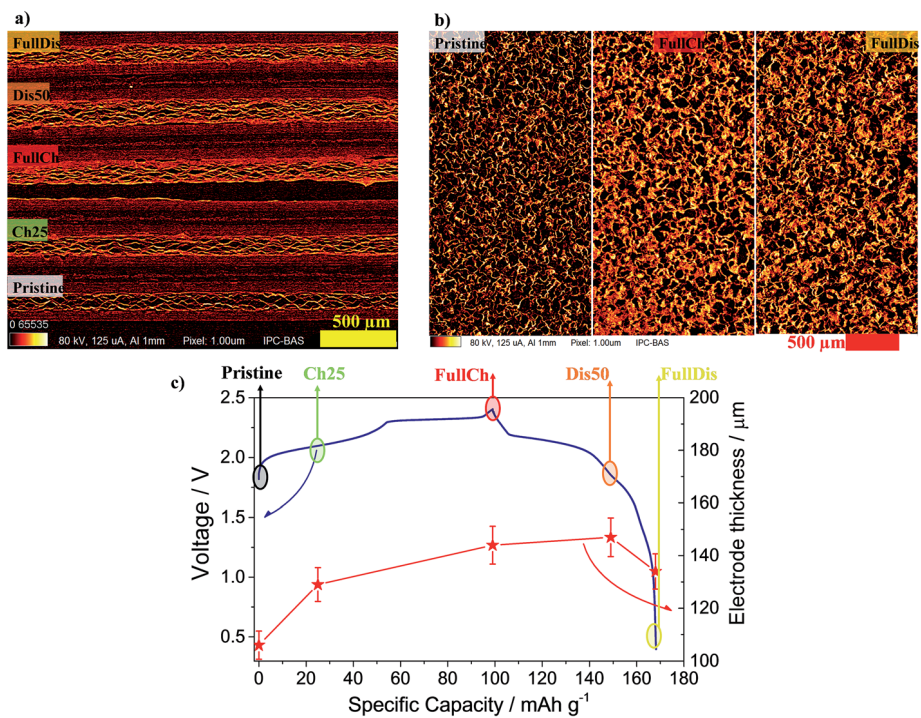

**Fig. 2** (a) CT cross-sections of four charged electrodes and pristine PG. (b) In-plane cross-sections of three electrodes at different charge states. (c) The red stars indicate the average thickness of the electrodes with various degrees of intercalation (Pristine, 25Ch, FullCh, 50Dis, FullDis) in relation to the voltage profile of the Al/EMIMCl:AlCl$_3$/PG cell galvanostatically cycled at 25 mA g$^{-1}$ current at 25 °C in the 0.4 to 2.4 V cut-off voltage range.

grayscale index represents the graphite or graphite plus the intercalated products.

The histograms shown in Fig. 3 are normalized to the area of the peak of the voids, and shifted to have the maximum of the peak of the voids equal to its average from all five curves. The obvious result is the relative increase of the peak of the graphite-based phase during charging followed by a mild decrease during discharging. The area below each peak divided by the sum of both peak areas gives the volume fraction of the corresponding phase. Thus, Fig. 4 represents the volume fraction of the voids, *i.e.* the porosity of the electrodes. The inset of Fig. 3 depicts the positions of the peaks of the graphite-based phase, *i.e.* carbon plus intercalated products.

Looking at Fig. 4, the two methods of determination of the porosity show the same trend during charging and discharging, but different values. That is expected, since the segmentation method contains an error equal to the difference between the overlapping parts of the two Gaussian distributions fitting the histogram. The histogram analysis itself has an error associated with fitting the data, but in our case should be considered to be more accurate. The porosity data obtained by segmentation are consistent with the structure separation data since they are an output of the same calculation procedure. However, using the Panasonic data for a PG density of 0.85 g cm$^{-3}$ and a density of graphite of 2.1 to 2.26 g cm$^{-3}$,[54] we obtain a porosity of the pristine PG of 60 to 62%, which is closer to that obtained by segmentation (66%) than that by histogram analysis (70%). Moreover, the histogram analysis also allows determination of the peak shifts that correspond to changes in X-ray absorption. As the X-ray absorption is correlated with (electron) density, the inset in Fig. 3 tells us that the mass density of the carbon-based phase increases during the intercalation of $AlCl_4^-$. Since Al and Cl contain more electrons than carbon, this demonstrates a homogeneous and bulk intercalation process. It should also be noted that the density does not fully restore after discharge, *i.e.* some of the $AlCl_4^-$ ions remain trapped in the electrode. The segmentation analysis also indicates that the entire porosity reported in Fig. 4 is open. For all samples reported, the closed porosity is less than 0.5% and may be considered insignificant. As far as the PG structure is layered, the pore structure is also layered in the smallest dimension, *i.e.* the layer thickness, perpendicular to the electrode plane. The structure separation can be considered as an average estimate of this dimension. As shown in Fig. S3,† the structure separation, *i.e.* the pore thickness, decreases from 9.5 μm to 8.5 μm during charging.

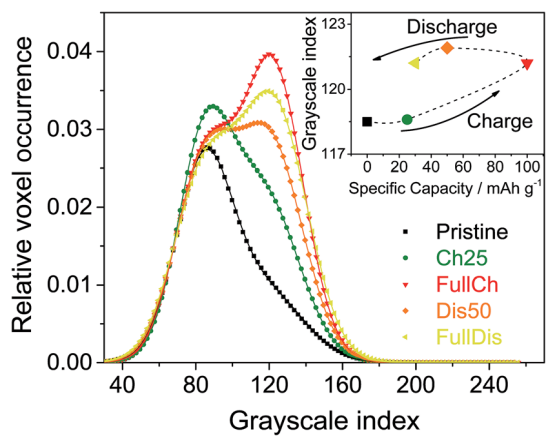

**Fig. 3** Histograms of the 3D images of PG electrodes at different charge states. The inset shows the peak positions for the carbon-based phase.

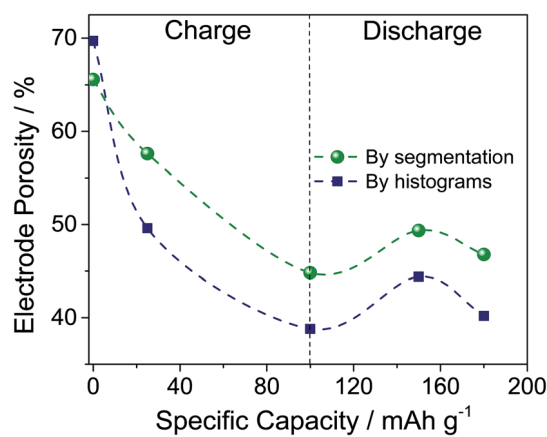

**Fig. 4** Porosity of the carbon electrodes (dashed lines are a guide for the eyes only).

Upon charging, the expansion of the carbon matrix due to intercalation results in an increase of the electrode thickness and a decrease of its porosity. The structure thickness associated with the thickness of the graphite layers increases from 6 to 9 μm, which correlates well with the increase of the electrode thickness from 100 μm to approximately 150 μm (*i.e.* 50% increase). Since the *c*-axis of the graphite grains in PG is oriented perpendicular to the sample plane, the sample thickness, $t$, can be easily calculated using eqn (4):[51]

$$t = t_0 \left( 1 + \frac{\Delta d}{0.335 n} \right). \tag{4}$$

where $t_0 = 100$ μm and $\Delta d = 0.59$ nm is the average of the obtained $\Delta d$ from SAXS, as reported in Table S3.† The theoretical line in Fig. 5 is calculated according to this equation with parameter values as mentioned above. As seen in Fig. 5 and also in Table S4,† the calculated thickness using the dominant stage determined by SAXS matches the measured thickness very well. XRD and SAXS data are consistent with the sample thickness determined using CT.

It was observed using all three techniques that during discharge the electrode does not return to its initial state. The final state, as determined from porosity, structure thickness, structure separation, and X-ray absorption, is closer to the charged state, rather than to the initial pristine PG. The electrode remains expanded, the porosity decreased slightly, and the size and density of the graphite-based phase increased. Indeed upon discharge the electrode thickness does not reach its initial value, remaining at 134 μm (Fig. 2c) due to the partial retention of anions intercalated in the graphite and to the incomplete reversibility of the electrochemical process, in agreement with the results obtained using XRD and SAXS.[25]

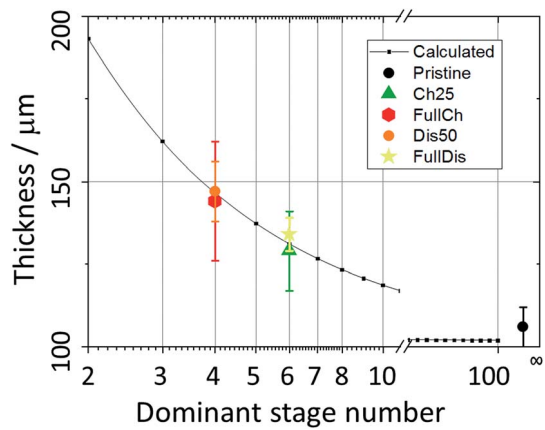

**Fig. 5** The PG electrode thickness as a function of the dominant intercalation stage.

The large decrease in the porosity leads to a limitation in the wettability of the electrode, most likely resulting in the formation of inactive areas in the electrode which could contribute to the first cycle irreversible capacity. The observed decrease of the electrode porosity that is most likely associated with the partial retention of $AlCl_4^-$ after the first cycle can be associated with a non-homogeneous electrode expansion in which some parts of the electrode expand and shrink earlier than the others. In contrast a homogeneous expansion of the electrode would preserve the porosity. An indication in favor of this hypothesis is the observation of coexisting intercalation phases with different intercalation stages, as in the Dis50 sample. Thus, the inhomogeneous expansion causes internal stress that is relieved by relative movements of the crystal grains leading to a decrease of the porosity. Since the intercalating species are not expected to cross the graphite basal planes,[51,55] the high orientation of the PG is unfavorable for the intercalation. The observed structural changes on a micro-level lead to relief from this unfavorable situation in terms of cycle efficiency, but at the expense of some of the $AlCl_4^-$ ions being trapped in the electrode. It should also be noted that the staging of the intercalation is driven by elastic forces[51] and an inhomogeneous distribution of mechanical stress due to inhomogeneous expansion could lead to irregularities in the intercalation stages. Generation and release of mechanical stresses is peculiar to PG since the crystallites contact each other on crystallographic planes parallel and perpendicular to their (001) axis, contrary to natural graphite powder, where the crystallites contact each other randomly. Therefore, the probability of occasional covalent bonding between the crystallites is much higher in the case of PG than in natural graphite. No stresses would be generated in a natural graphite electrode and consequently no irreversibility in the first cycle would be observed. The latter has been recently confirmed, using graphite powder as an electrode, where a coulombic efficiency close to 100% has been obtained in the first cycle.[23]

## Conclusions

In this work a pyrolytic graphite cathode material for aluminum batteries has been investigated by combining SAXS, XRD and CT. A good correlation between the periodic repeat distances as a function of intercalation was obtained, demonstrating that the intercalation process involves the formation of a stage 4 GIC in the fully charged state. The morphological evolution of the electrode upon cycling was investigated using CT measurements. An increase in the periodic repeat distance of the GIC nanostructure upon charging was directly correlated to the expansion of the graphite electrode microstructure.

The volumetric expansion leads to a very large decrease of the electrode porosity, and a reduction of the pore size. This phenomenon was indicated as the most probable reason for the first cycle irreversible capacity. The inhomogeneous stresses created and relieved during the first intercalation are irreversible and therefore responsible for the trapping of a fraction of the intercalated ions. This behavior is limited to PG due to the orientation of its crystal grains around the (001) axis. Partial isolation of electrode areas and their limited electrochemical activity due to microstructural changes associated with the decrease in porosity cannot be excluded.

In conclusion, the combination of SAXS, XRD, and CT gives us a complete overview of the micro/nanostructure of the studied system, and allows us to establish a powerful method suitable for the structural refinement of energy-related systems.

## Conflicts of interest

There are no conflicts to declare.

## Acknowledgements

This research was funded by the European Commission in the H2020 ALION project under contract 646286 and the German Federal Ministry of Education and Research in the AlSiBat project under contract 03SF0486. We acknowledge the X-ray CoreLab at Helmholtz-Zentrum Berlin for allocation of X-ray powder diffraction time and support. In particular, we would like to thank Dr Michael Tovar and Mr Rene Gunder for their kind availability.